\newcommand{\tindex}[1]{{\scriptstyle {\rm#1}}}
\documentstyle[pre,aps,epsf,floats]{revtex} 

\begin{document} 
\flushbottom
\draft

\twocolumn[\hsize\textwidth\columnwidth\hsize\csname @twocolumnfalse\endcsname

\title{
  Creep motion in a random-field Ising model
  } 
\author{
  L.~Roters\cite{Lars_seine_Email},
  S.~L\"ubeck\cite{Sven_seine_Email},
  and K.\,D.\ Usadel\cite{Usadel_seine_Email}
  }
\address{
  Theoretische Tieftemperaturphysik, \\
  Gerhard-Mercator-Universit\"at Duisburg,\\ 
  47048 Duisburg, Germany \\
  }
\date\today
\maketitle
\begin{abstract}
  We analyze numerically a moving interface in the
  random-field Ising model which is driven by a magnetic field. 
  Without thermal fluctuations the system displays
  a depinning phase transition, i.e., the interface is pinned
  below a certain critical value of the driving field.
  For finite temperatures the interface moves even 
  for driving fields below the critical value. 
  In this so-called creep regime the
  dependence of the interface velocity on the temperature is 
  expected to obey an Arrhenius law.
  We investigate the details of this Arrhenius behavior in 
  two and three dimensions and compare our 
  results with predictions obtained from renormalization 
  group approaches. 
\end{abstract}
]
\narrowtext

\section{Introduction}
\label{intro}
\begin{figure}[t]
 \epsfxsize=8.0cm
 \epsffile{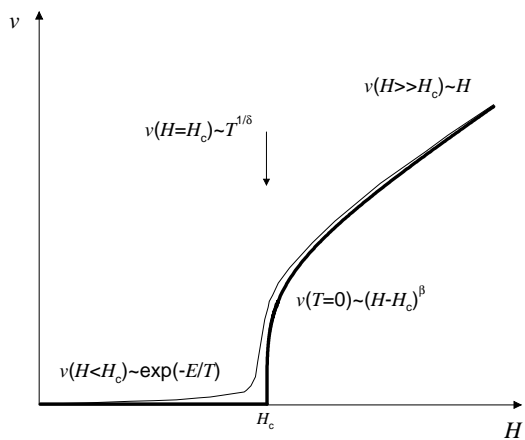}
 \caption{Schematic sketch of the interface velocity $v$ of the 
          pinning phase transition and its dependence on the driving
          field $H$. 
          The bold line corresponds to zero temperature, $T=0$.
          For small but finite temperatures 
          the critical behavior is smeared out (thin solid line).
          The creep regime for small driving fields
          is characterized by an Arrhenius like
          behavior with an effective energy barrier $E$. 
 \label{fig:phase_dia_01}} 
\end{figure}

In recent years the understanding of driven interfaces has 
improved considerably. 
Well known models of such interfaces are the equations
of Edwards and Wilkinson~\cite{EW}
as well as of Kardar, Parisi, and Zhang~\cite{KPZ}.
Of particular interest are driven interfaces moving through a 
quenched disordered medium  which exhibit a so-called
depinning phase transition.
Without disorder the velocity of a driven interface grows linearly 
with the applied driving force or driving field, respectively.
This behavior changes in the presence of quenched disorder.
For small driving fields the interface is pinned
by the disorder.
The interface moves only if the driving field exceeds
a critical value, i.e., on increasing the driving field
a continuous phase transition from a pinned to a 
moving interface takes place
(see, for instance,~\cite{LNST_REV} and references therein).
The expected dependence of the interface velocity on the
driving field is sketched in Fig.~\ref{fig:phase_dia_01}.
For very large driving fields the disorder can be
neglected and consequently the velocity depends 
linearly on the driving field.
The depinning transition happens due to the competition
between the disorder and the driving field.
The disorder induces some effective energy barriers
that  suppress the interface motion.
The driving field reduces these energy barriers 
but they are overcome only if the driving field
exceeds the critical value.
Examples of systems exhibiting a depinning transition are 
charge density waves~\cite{CHARGES1,CHARGES2}, 
or field driven domain walls in ferromagnets~\cite{BRUINSMA1}. 

However, in the above scenarios thermal fluctuations
are neglected.
In real systems these fluctuations occur and 
no critical behavior is observed for finite temperatures.
The reason is that even below the critical driving field
the energy barriers can be
overcome due to thermal fluctuations, 
resulting in a moving interface.
A striking effect of thermal fluctuations 
occurs at the critical field where the 
interface velocity $v$ depends on the temperature $T$ according
to $v\sim T^{1/\delta}$ 
with an exponent
$\delta \geq 1$~\cite{CHARGES2,MIDDLETON1,NOWAK1,ROTERS1}. 
Another effect of thermal fluctuations
is the so-called creep motion that occurs for  
driving fields $H$ well below the critical threshold
at sufficiently low temperatures.
Here, the interface velocity is expected to be
characterized by an Arrhenius behavior
\begin{equation}
  v \sim {\rm e}^{-E(H)/T}
  \label{eq:exp}
\end{equation}
with a certain field dependent energy barrier $E(H)$.
Creep motion was investigated, for instance, within the theory of
flux creep phenomena~\cite{FEIGELMAN1} and in several renormalization
group approaches to the Edwards-Wilkinson
equation~\cite{CHAUVE1,NATTERMANN2}.
Experimentally, the Arrhenius behavior of the creep regime
was observed for magnetic 
domain wall motion in thin films composed 
of Co and Pt layers~\cite{FERRE1}.

In this paper we consider the interface motion occurring in a
driven random-field Ising model (RFIM) in the creep
regime.
In the next section we describe the details of the model and
the simulations.
In Secs.~\ref{D2} and~\ref{D3} we investigate numerically the
creep motion of the interface in the two and three 
dimensional RFIM. 
In particular, we show that the velocity behavior
can be described by an Arrhenius ansatz and we investigate
the temperature and field
dependence of the prefactor of the
Arrhenius law and study the energy barrier~$E(H)$. 
In section~\ref{dis} we summarize and discuss our results.


\section{Model and Simulations}
\label{model}
To study the creep regime we consider the RFIM
on a square or simple cubic lattice of linear size $L$.
The Hamiltonian of the RFIM is given by
\begin{equation}
  {\cal H} =
  -\frac{J}{2}\, \sum_{\langle i,j \rangle} S_i \, S_j
  -H \, \sum_{i} S_i
  - \sum_{i} h_i\,S_i
  \mbox{ ,}
\end{equation}
where the first term characterizes the exchange interaction 
of neighboring spins ($S_i=\pm 1$).
The sum is taken over all pairs of nearest neighbor spins.
The spins are coupled to a homogeneous driving field $H$ and to 
quenched random-fields $h_i$ which we choose to be uncorrelated 
($\langle h_i h_j \rangle \propto \delta_{ij}$)
with $\langle h_i \rangle = 0$. 
Throughout this paper we consider uniformly distributed
disorder, i.e., the probability density $p$ that the random field takes
some value $h_i$ is given by
\begin{equation}
  p(h_i) = 
  \left\{
    \begin{array}{ccl}
      (2\Delta)^{-1} & \;{\rm for}\; &|h_i| < \Delta\\
      0 & & \mbox{otherwise.}
    \end{array}
  \right.
\end{equation}
Using antiperiodic boundary conditions an
interface is induced in the system
which can be driven by the field $H$
(see~\cite{ROTERS1} for details).
Starting with an initially flat interface we apply
a Glauber dynamics with random sequential 
update and heat-bath transition probabilities
(see, for instance,~\cite{BINDER_BOOK}). 
In our simulations the interface moves along the
[11]- and [111]-direction of a simple cubic lattice.
This is a natural choice since in the absence of disorder
interface motion occurs for any finite driving 
field~\cite{NOWAK1}. 
This property is an advantage especially in the creep regime
where the interface is driven by small 
driving fields at low temperatures.

The basic quantity in our simulations is the 
velocity of the moving interface, which is determined
in the following way.
The interface movement corresponds to an increasing
magnetization which is monitored as a function
of time, i.e., the number of Monte Carlo steps per spin. 
Starting from a flat interface the system 
after a certain transient regime reaches a steady
state where the average magnetization grows linearly
in time.
The velocity of the interface is defined
as the time derivative of this magnetization.
Spin flips outside the interface may also occur,
caused by the finite temperature.
These isolated, rare spin flips are unstable 
for sufficiently small temperatures, i.e., 
they flip back in the next update.
Thus these spin flips do not affect the measurement
of the global average magnetization time dependence
and therefore do not affect the determination of the interface
velocity. 

During its motion the originally flat interface roughens due to the
disorder. The width of the interface increases and finally reaches a
stationary state. For the data presented in this paper we have
verified that the interface width remains small as compared to
the extension of the system perpendicular to the interface. 


\section{Creep motion in the two dimensional RFIM model}
\label{D2}

We measured the
velocity of the interface in the creep motion regime.
Since the creep regime is ``far away'' from the critical
point, we expect that finite-size effects can be neglected.
Investigations of various system sizes confirm
this assumption and we use therefore in our
simulations a large number of update steps instead of large
system sizes.
We performed for each temperature and field value 
at least $10^5$ Monte Carlo steps. 
Additionally we focus our analysis 
on one value of the disorder strength ($\Delta=1.2$).
We have also performed some spot checks
at different values of $\Delta$ in order to confirm
that the results are not sensitive to the disorder strength.

\begin{figure}[b]
 \epsfxsize=8.0cm
 \epsffile{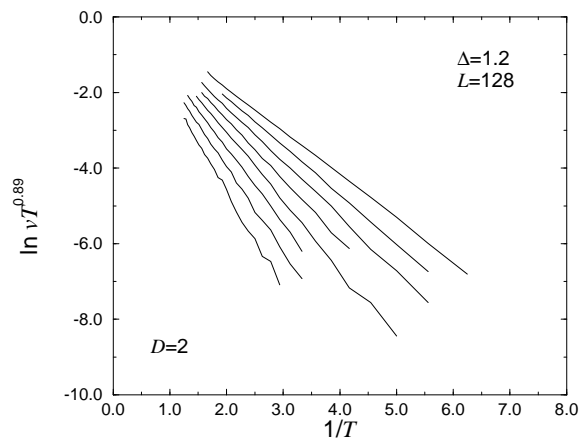}
 \caption{The interface velocity~$v$ as a function of
          the temperature~$T$ for various values of the driving 
          field ($H\in\{0.25, 0.3, 0.35,...,0.6\}$ from bottom to
          top) and $\Delta=1.2$.
          According to Eq.~(\protect\ref{eq:v_creep_01})
          we plot $\ln{v T^{x}}$ vs~$1/T$.
          On varying the exponent~$x$ we obtained nearly straight
          lines for $x=0.89\pm 0.17$.
          The cutoffs at low and high temperatures are caused
          by different effects.
          At low temperatures the interface can be pinned for
          finite time intervals depending on the particular
          disorder configuration.
          In this case the interface displays a stop and go
          behavior which results in strong velocity fluctuations
          (not shown for clarity).
          The cutoff for large $v$ occurs because
          the creep regime is exited at
          high temperatures or high driving fields.
 \label{fig:rfim_2d_arrh_01}
 } 
\end{figure}
As mentioned above the velocity is expected to obey 
an Arrhenius law
\begin{equation}
v(H,T) \; = \; C(H,T) \; e^{-E(H)/T}
\label{eq:v_creep_01}
\end{equation}
in the creep regime.
The effective energy barrier~$E(H)$ is independent of the 
temperature and tends to zero for $H \to H_{\rm c}$.
Following a renormalization
group analysis~\cite{CHAUVE1} we assume
that the temperature dependence 
of the prefactor of the Arrhenius law is
characterized by a power-law behavior
\begin{equation}
C(H,T) \; = \; c(H) \, T^{-x}
\label{eq:prefactor}
\end{equation}
with some particular exponent~$x$.
Independent of the actual value of  $x$ the interface
motion stops for any finite value of the energy barrier
($H<H_\tindex{c}$) in the limit $T \to 0$.

In the first step of our analysis we
determine the exponent~$x$.
In an Arrhenius plot $\ln{v T^x}$ vs~$1/T$ 
the exponent~$x$ is varied until 
straight lines are obtained.
Good results are found for $x =0.89 \pm 0.17$
and the corresponding curves are shown 
in Fig.~\ref{fig:rfim_2d_arrh_01}.

\begin{figure}[t]
 \epsfxsize=8.0cm
 \epsffile{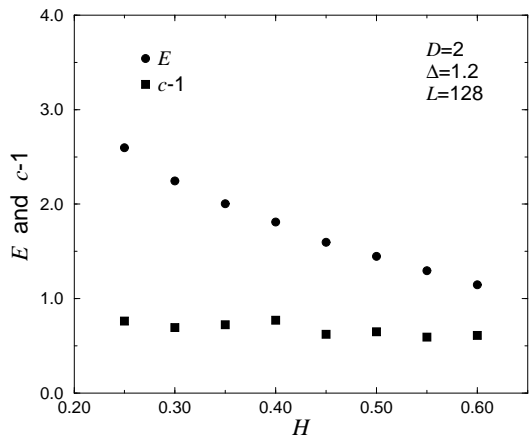}
 \caption{The energy barrier $E$ and the prefactor $c$ 
          [see Eqs.~(\protect\ref{eq:v_creep_01}) and
          (\protect\ref{eq:prefactor})]
          as a function of the driving field~$H$.
          To avoid an overlap between the two curves we plot
          $c-1$ instead of $c$.
 \label{fig:reg_2d_E_c_01}} 
\end{figure}

\begin{figure}[t]
 \epsfxsize=8.0cm
 \epsffile{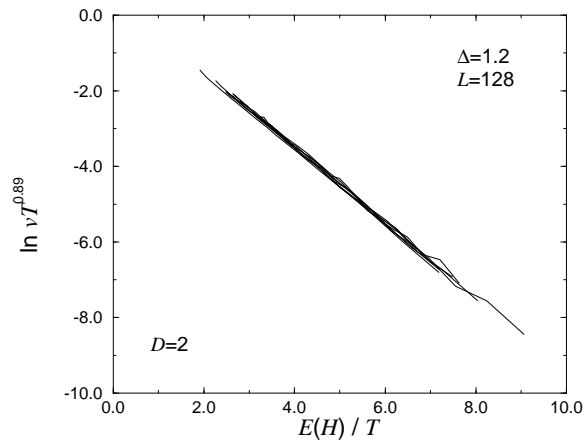}
 \caption{The rescaled interface velocity $\ln{v T^{x}}$
          vs~$E(H)/T$.
          The values of the energy barrier $E(H)$ are obtained
          from a regression analysis of the corresponding
          curves of Fig.~\protect\ref{fig:rfim_2d_arrh_01}.
          The data collapse confirms that the prefactor
          $c$ displays no significant field dependence
          [Eqs.~(\protect\ref{eq:v_creep_01}) and
          (\protect\ref{eq:prefactor})].
 \label{fig:rfim_2d_arrh_02}} 
\end{figure}

A regression analysis of these curves then yields 
the value of the prefactor $c(H)$ and the value of the energy
barrier $E(H)$ 
[Eqs.~(\ref{eq:v_creep_01}) and (\ref{eq:prefactor})].
The results are plotted in Fig.~\ref{fig:reg_2d_E_c_01}.
On increasing the driving field the effective 
energy barrier decreases as expected.
But the prefactor of the Arrhenius law displays no 
significant field dependence, i.e., $c(H)={\rm const}$.
This is confirmed by Fig.~\ref{fig:rfim_2d_arrh_02}
which was obtained by plotting $\ln{v T^x}$ as 
a function of $E(H)/T$. 
The curves for different values of the driving field are seen to
coalesce to a single curve which happens only if $c(H)={\rm const}$. 

\begin{figure}[t]
 \epsfxsize=8.0cm
 \epsffile{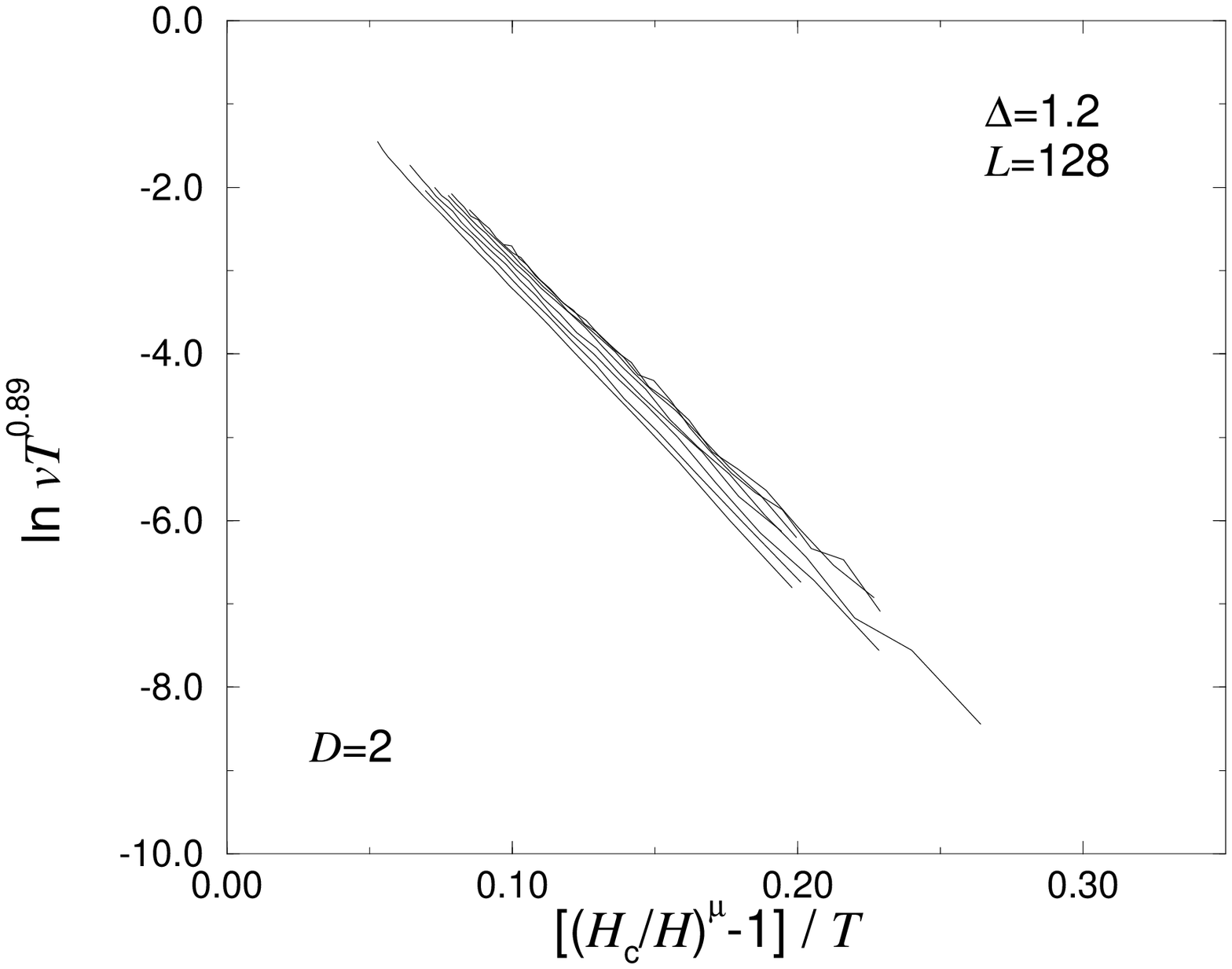}
 \caption{The rescaled interface velocity $\ln{v T^{x}}$
          as a function of $[(H_{\rm c}/H)^{\mu}-1]/T$
          [see Eq.~(\protect\ref{eq:E_ansatz_power_01})].
          No coalescence of the data could be obtained for any finite
          value of~$\mu$.
          But with decreasing exponent $\mu$ the coalescence becomes
          better. 
          The figure shows the corresponding curves for $\mu= 0.05$.
 \label{fig:rfim_2d_arrh_03}} 
 \epsfxsize=8.0cm
 \epsffile{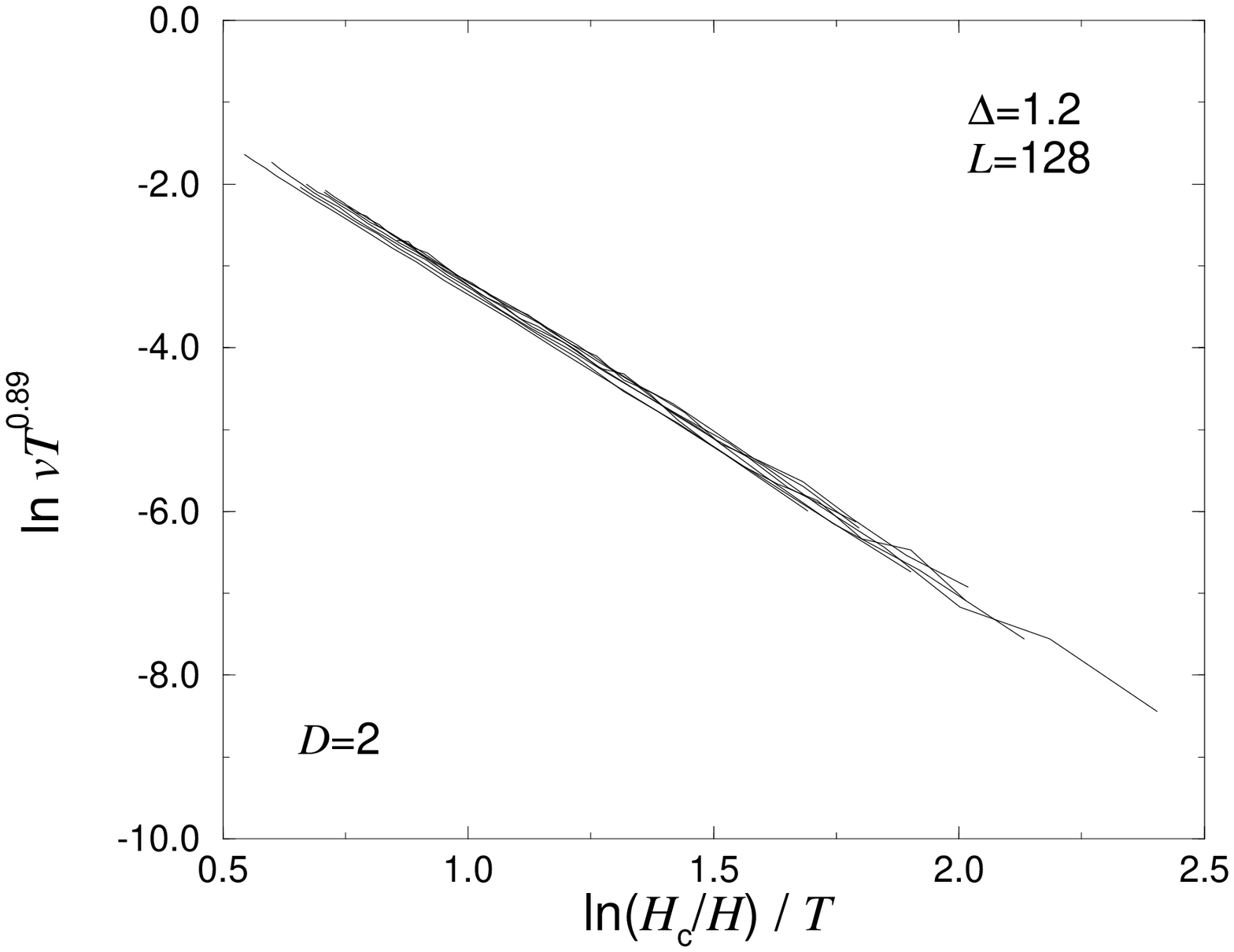}
 \caption{The rescaled interface velocity $\ln{v T^{x}}$ as a function
          of $\ln{(H_{\rm c}/H)}/T$. 
          This logarithmic {\it Ansatz} yields a quite convincing data
          collapse, i.e., this result suggests that the field
          dependence of the energy barrier is given by 
          Eq.~(\ref{eq:E_ansatz_power_01}) in the limit $\mu\to 0$.
 \label{fig:rfim_2d_arrh_05}} 
\end{figure}
We analyze the field dependence of the energy barrier
starting from a recently performed renormalization group
approach~\cite{CHAUVE1}, assuming
\begin{equation}
E(H) \; = \; E_{\scriptscriptstyle 0} \,
\left [ \left(\frac{H_{\rm c}}{H} \right )^\mu -1\right ].
\label{eq:E_ansatz_power_01}
\end{equation}
On approaching the critical field of the depinning
transition $H_{\rm c}$ the energy barrier vanishes.
The value of the critical field $H_{\rm c} = 1.12\pm 0.02 $ is 
obtained from an independent simulation at zero 
temperature.
Thus we plotted the rescaled velocities as a function
of $(H_{\rm c}/H)^\mu -1$ and tried to obtain a 
coalescence of the data similar to
Fig.~\ref{fig:rfim_2d_arrh_02} by varying the exponent~$\mu$.
(See Fig.\ref{fig:rfim_2d_arrh_03}).
Our analysis shows that only a logarithmic field dependence for
$\mu \to 0$ fits the data, i.e., 
\begin{equation}
E(H) \; = \; E_{\scriptscriptstyle 0}\, 
\ln{\left ( \frac{H_{\rm _c}}{H} \right)}
\label{eq:E_ansatz_log}
\end{equation}
(see Fig.~\ref{fig:rfim_2d_arrh_05}).
As one can see in Fig.~\ref{fig:rfim_2d_arrh_05} the logarithmic
{\it Ansatz} yields a quite convincing fit.
Our analysis therefore suggests that the effective energy 
barrier displays a logarithmic field dependence.


\section{the three dimensional model}
\label{D3}

\begin{figure}[t]
 \epsfxsize=8.0cm
 \epsffile{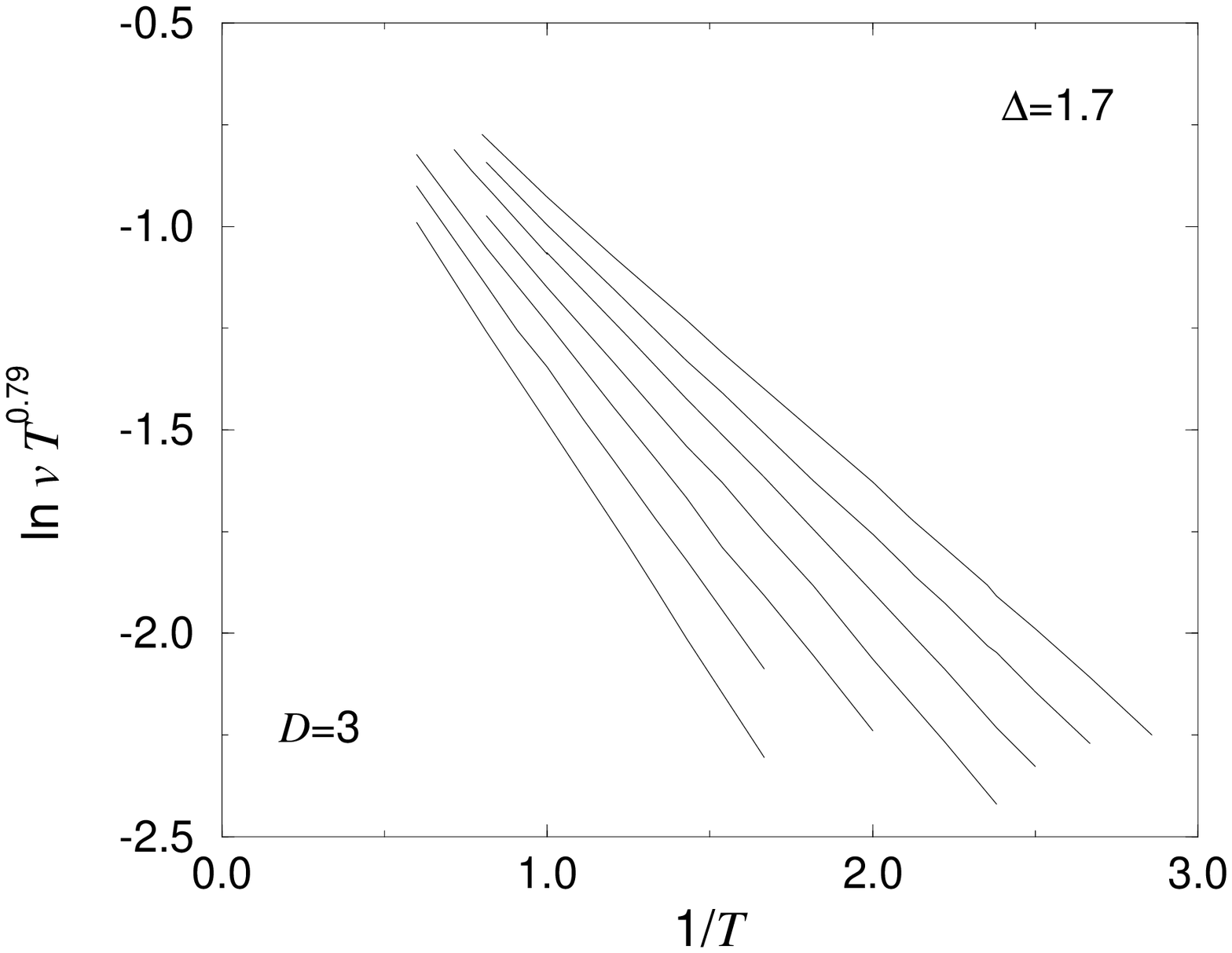}
 \caption{
   Interface velocities of the three dimensional  
   model for different temperatures and
   driving fields ($H=0.3,0.35,0.4,...,0.6$, from bottom to top). 
   On varying the exponent~$x$ we obtained nearly straight lines
   for $x=0.79\pm 0.09$.
   \label{fig:d3_vels}
   } 
 \epsfxsize=8.0cm
 \epsffile{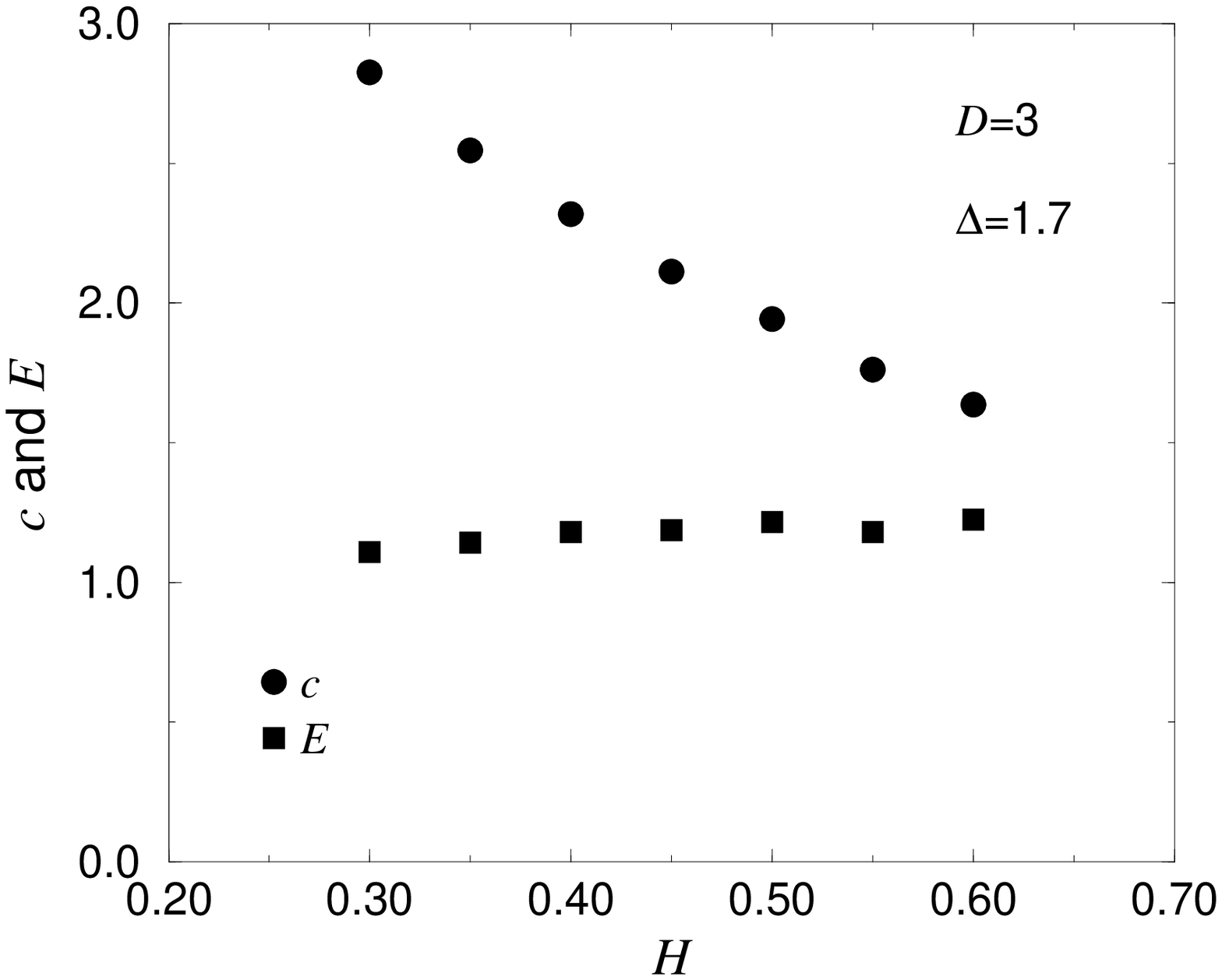}
 \caption{
   The energy barrier $E$ and the prefactor $c$ 
   [see Eq.~(\ref{eq:prefactor})] as a function of the 
   driving field $H$. 
   \label{fig:d3_ec}
   } 
\end{figure}
We next analyze the interface velocity of the three dimensional model
for $\Delta=1.7$. 
For this value of the disorder the critical behavior has been
investigated and the corresponding critical field has been found to be
$H_{\rm c}=1.37\pm 0.01$~\cite{ROTERS1}.
As for the two dimensional model we have also performed some
simulations for values different from $\Delta = 1.7$ to ensure 
that the main results discussed below do not depend on 
the particular choice of $\Delta$.

By driving the interface at finite temperatures 
and fields below the critical threshold
we measured the velocity $v(H,T)$.
Again, we fitted the simulation data to
Eqs.~(\ref{eq:v_creep_01}) and (\ref{eq:prefactor})
by varying  $x$ to get straight lines in the
$\ln{v T^x}$ vs~$1/T$ plot.
A good fit is obtained 
using $x=0.79\pm 0.09$~(Fig.~\ref{fig:d3_vels}).

\begin{figure}
 \epsfxsize=8.0cm
 \epsffile{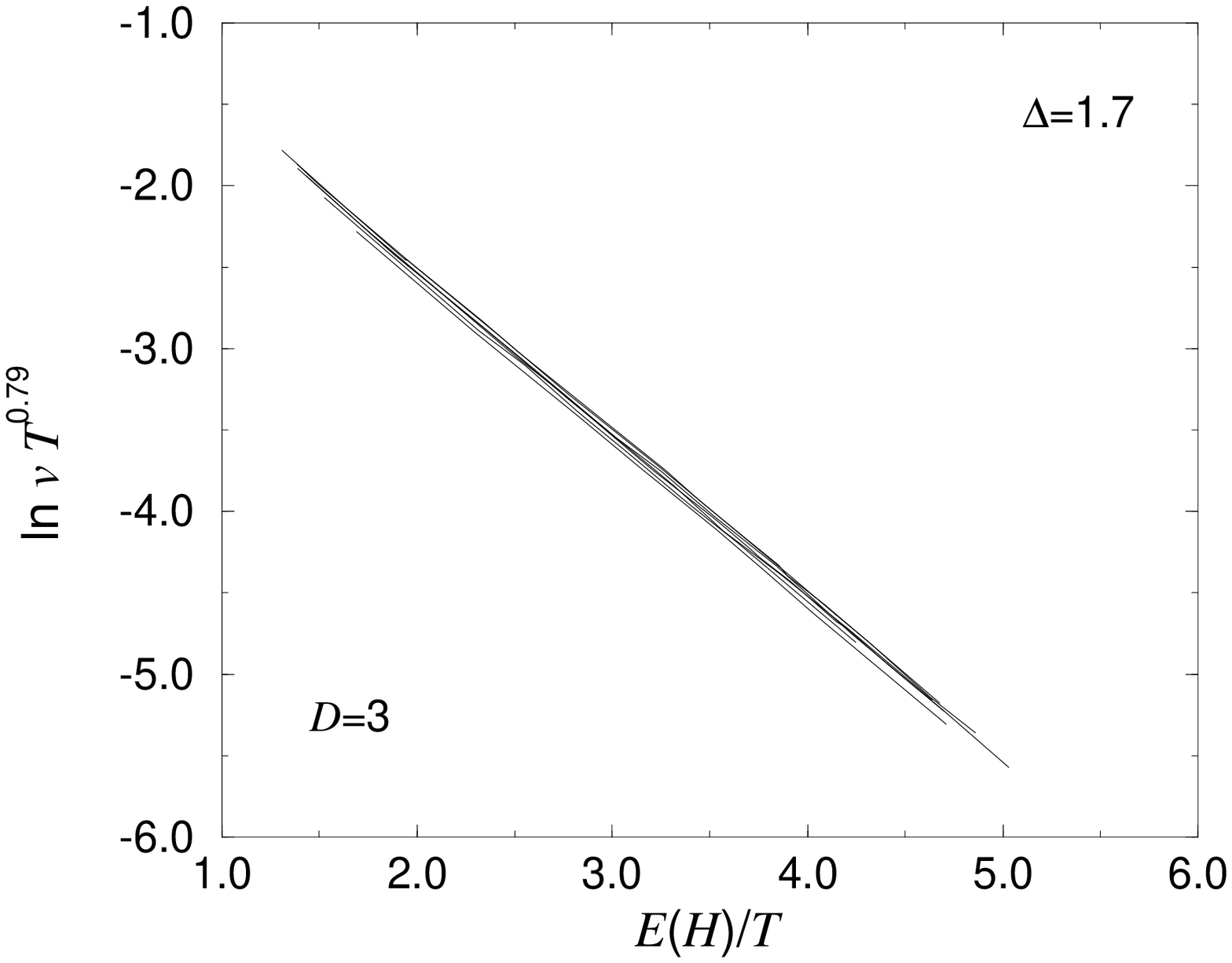}
 \caption{
   On rescaling the interface velocities with the numerically
   determined energy barrier $E(H)$ the data shown in
   Fig.~\ref{fig:d3_vels} coalesce onto one single curve. 
   As in the two dimensional case this behavior shows that
   $c(H) \approx \mbox{const}$ [see Eq.~(\ref{eq:prefactor})]. 
   \label{fig:e-sim}
   } 
\end{figure}

The result of the regression analysis for $E(H)$ and $c(H)$ is shown
in Fig.~\ref{fig:d3_ec}. 
As in the two dimensional case $E(H)$ decreases with
increasing driving field and the prefactor 
is essentially independent of the driving field.
On plotting the rescaled interface velocities vs~$E(H)/T$
the data coalesce  (see Fig.~\ref{fig:e-sim})
only if $c(H) ={\rm const}$. 

\begin{figure}[t]
 \epsfxsize=8.0cm
 \epsffile{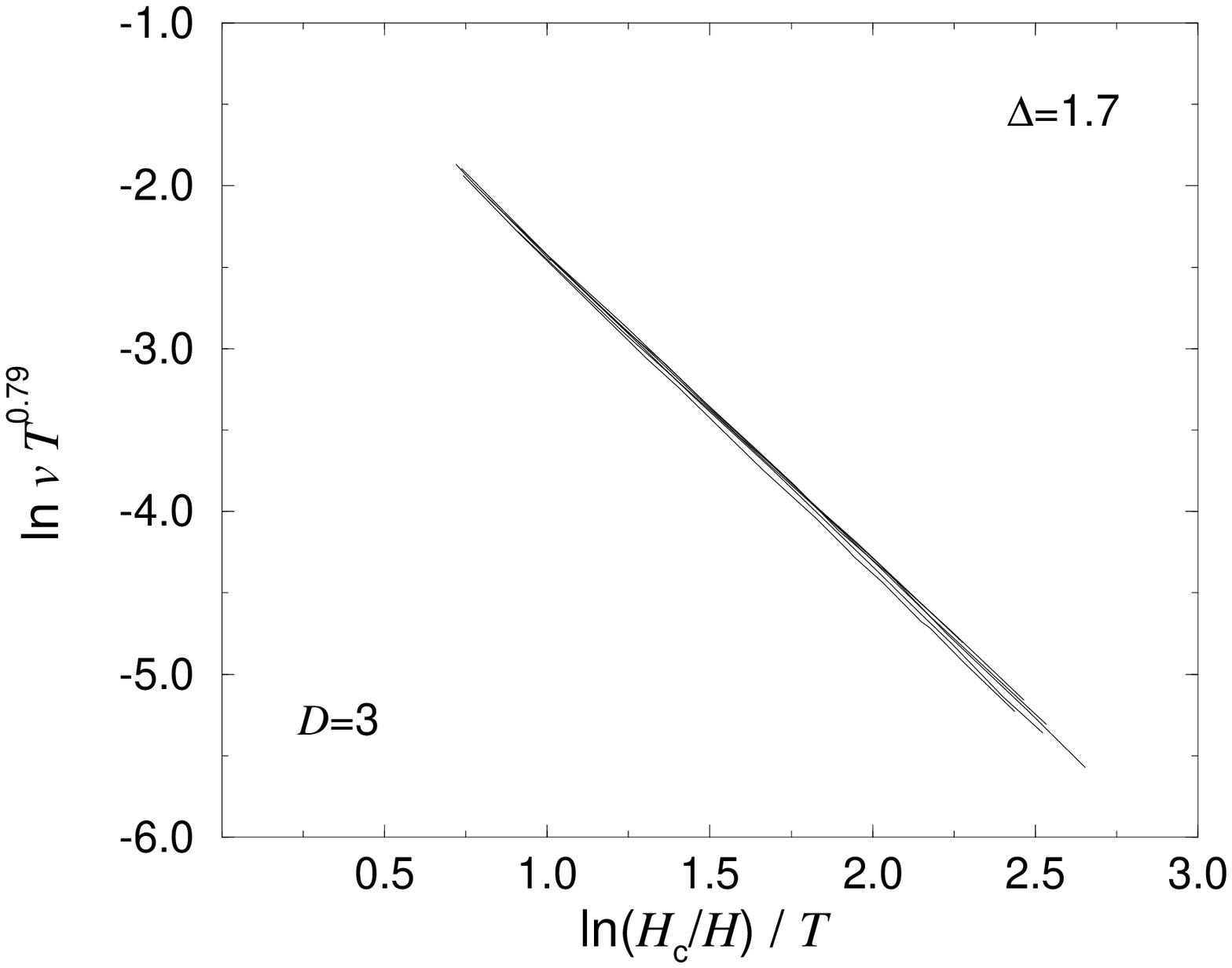}
 \caption{
  Equation~(\ref{eq:E_ansatz_power_01}) results in a logarithmic 
  dependence of the energy barrier on the driving field
  for $\mu \to 0$. The figure shows the interface velocities,
  which are rescaled according to Eq.~(\ref{eq:E_ansatz_log}). 
  \label{fig:e-log}
   } 
\end{figure}

We consider now the field dependence of the
energy barrier.
In analogy to the previous section we check
the conjectured field dependence obtained 
from a renormalization group approach.
Applying the data of the interface velocities to the {\it Ansatz} 
Eq.~(\ref{eq:E_ansatz_power_01}) yields a similar 
result as in two dimensions, i.e., the accuracy of the 
data collapse increases for $\mu\to 0$.  
Therefore we again assume that the dependence of the energy 
barrier on the driving field displays a logarithmic 
behavior [Eq.~(\ref{eq:E_ansatz_log})].
The corresponding curves are shown in
Fig.~\ref{fig:e-log}. 
As can be seen the logarithmic ansatz yields 
a good fit of the velocity data in the  
creep regime.

\begin{figure}[t]
 \epsfxsize=8.0cm
 \epsffile{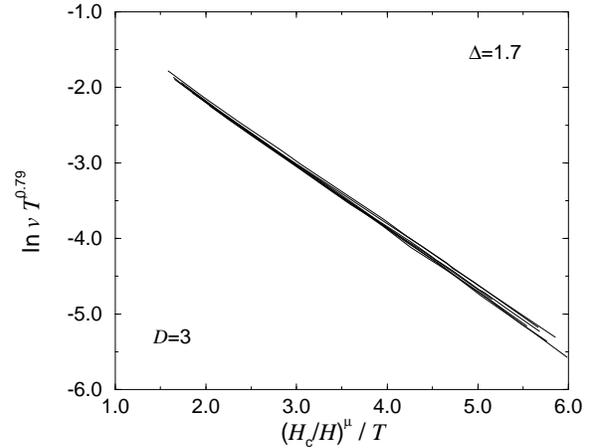}
 \caption{
   Scaling plot for the {\it Ansatz} of the energy barrier according
   to Eq.~(\ref{eq:E_ansatz_power_03}).
   From the data collapse one obtains $\mu=0.825 \pm 0.1$.
   } 
   \label{fig:e-ferre}
\end{figure}
But we have to admit that in contrast to the
two dimensional case we observe here that
different {\it Ans\"atze} for the field dependence 
of the energy barrier may also lead to
a fit of the data. 
For example, several authors
conjectured that the energy barrier is given by
\begin{equation}
  E(H)  =  E_{\scriptscriptstyle 0}  
     \left( \frac{H_\tindex{c}}{H} \right) ^\mu
  \label{eq:E_ansatz_power_03}
\end{equation}
for $H \ll H_\tindex{c}$.
Note that Eq.~(\ref{eq:E_ansatz_power_03})
agrees with the one discussed above
[Eq.~(\ref{eq:E_ansatz_power_01})] for sufficiently 
small driving fields.
The above {\it Ansatz} was derived within a theory of
flux creep behavior~\cite{FEIGELMAN1} and is expected to 
hold for the present situation of driven interfaces.
In this case the exponent $\mu$ is given by
$\mu = (2 \zeta + D - 3) / (2 - \zeta)$ with $\zeta$ denoting
the roughness exponent of the interface at the depinning transition
(see~\cite{NATTERMANN2,FERRE1} and references therein). 
For the Edwards-Wilkinson equation with quenched disorder, 
$\zeta = (5-D) / 3$ has been determined by an $\epsilon$ expansion
within a renormalization group scheme~\cite{NATTERMANN1}. 
This value is believed to be exact to all orders of
$\epsilon$~\cite{FISHER1}, and inserting
it into the formula above yields $\mu = 1$, independent
of $D$. 

Fitting our data according to Eq.~(\ref{eq:E_ansatz_power_03})
yields $\mu = 0.825 \pm 0.1$
(Fig.~\ref{fig:e-ferre}). 
The accuracy of the fit is similar to
the one obtained from the logarithmic
{\it Ansatz} of $E(H)$.
Thus, in the three dimensional case one cannot infer the correct
expression of the energy barrier from the accuracy of the data fit. 
On the other hand, the 
driving fields considered ($H=0.3,...,0.6$) are of the same order 
as the critical value ($H_{\rm c}\approx 1.37$)
while Eq.~(\ref{eq:E_ansatz_power_03})
is believed to be valid only in the 
limit $H\ll H_{\rm c}$.

\section{Discussion and conclusion}
\label{dis}

We investigated numerically the creep motion of a driven interface
of a RFIM model in the limit of low temperatures
and small driving fields.
We found that the interface velocity obeys an Arrhenius law, 
which was investigated in detail.
We assumed that the prefactor of the Arrhenius law can be written as 
$C(H,T) \sim c(H) \, T^{-x}$. 
Applying this {\it Ansatz} to the numerically determined interface
velocities, we find a positive exponent~$x$ for the two and three
dimensional model.
Additionally, our results suggest that $c(H)$ is independent
of the driving field in both cases.
These results are in contradiction to a renormalization group
analysis~\cite{CHAUVE1} in which (i) $x$ is claimed to be negative and
(ii) $c(H)$ is found to exhibit a significant field dependence 
$[c(H) \sim H^\sigma$ with $\sigma > 0]$.
In particular the opposite sign of the exponent $x$ is 
remarkable.

Knowing the prefactor $C(H,T)$, it is possible to investigate the
driving field dependence of the energy barrier $E(H)$.
Our numerical results are in agreement with the assumption that the
energy barrier depends logarithmically in both dimensions on the
driving field. 
Again this result is in contradiction with both phenomenological
theories and renormalization group approaches, which conjecture an
algebraic behavior~\cite{CHAUVE1,NATTERMANN2,FERRE1}.
The logarithmic behavior can be explained if one assumes 
that the exponent of the algebraic behavior tends to zero. 
But in~\cite{CHAUVE1,NATTERMANN2,FERRE1} a finite value
of the corresponding exponent is predicted.

Thus our analyses reveal that the driven interface of 
a RFIM displays creep motion in the limit of low temperatures
and small driving fields characterized by an Arrhenius 
law as predicted by phenomenological and renormalization group
approaches.
The details of the Arrhenius law (prefactor
and energy barrier) differ, however, from the predicted
behavior.
Further investigations are needed to understand
these differences.

\acknowledgments
This work was supported by the Deutsche Forschungsgemeinschaft via
Graduiertenkolleg {\it Struktur und Dynamik heterogener Systeme}
at the University of Duisburg, Germany.

\end{document}